\documentstyle[aps,12pt,epsfig]{revtex}

\parskip=\medskipamount

\def\be{\begin{equation}}
\def\ee{\end{equation}}
\def\disp{\displaystyle}

\begin{document}

\title{Adsorption of a random heteropolymer at a potential well revisited:
location of transition point and design of sequences}

\author{Alexei Naidenov$^{1}$ and Sergei Nechaev$^{1,2}$}

\address{$^{1}$L D Landau Institute for Theoretical Physics, 117940,
Moscow, Russia}

\address{$^{2}$UMR 8626, CNRS--Universit\'e Paris XI, LPTMS, Bat.100,
Universit\'e Paris Sud, \\ 91405 Orsay Cedex, France}

\maketitle

\begin{abstract}
The adsorption of an ideal heteropolymer loop at a potential point well is
investigated within the frameworks of a standard random matrix theory. On the
basis of semi--analytical/semi--numerical approach the histogram of transition
points for the ensemble of quenched heteropolymer structures with bimodal
symmetric distribution of types of chain's links is constructed. It is shown 
that the sequences having the transition points in the tail of the histogram 
display the correlations between nearest--neighbor monomers.
\end{abstract}

\section{Introduction}

The problem of adsorption of an ideal heteropolymer chain with random
quenched chemical structure (i.e. sequence of links) is far from being new
and is well represented in the literature.

The simple diffusive approach \cite{bin,bir2,bir3,gin} provides us complete
understanding of the homopolymer and block--copolymers adsorption in
various geometries. More advanced renormalization group methods
\cite{gro1,gro2} and power series analysis \cite{obu} applied to random
chains with a disordered sequence of links delivers an important
information about the thermodynamic properties of ideal polymers near the
transition point from delocalized (Gaussian) to localized (adsorbed)
regimes. A mathematical formalism based on perturbation theory gives
reasonable results for a phase transition in solid--on--solid (SOS) models
with quenched impurities \cite{forg}. Many conclusions of \cite{forg}
correlate with those obtained by RG analysis in \cite{gro2}. Problems
dealing with localization of disordered polymers at selective interfaces
\cite{orl,nezh,mon} should be also mentioned in the context of discussed
problems.

However even since early 80s during almost 20 years of intent attention
some important questions of heteropolymer adsorption remain still open. One
of them, the most intriguing, to our point of view, concerns the location
of a point of a phase transition from coil to localized (adsorbed at a
potential well) state. 

In this paper we consider the model of an ideal, i.e. self--intersecting {\it
ring} polymer chain consisting of two types of links, "black" and "white"
organized in different chemical structures having different energies in the
single point--like potential well. We distinguish between the following
models of chemical structures:
\begin{enumerate}
\item {\it "Canonically quenched"} sequence---positions of "black" and
"white" links are uniquely fixed for a given chain structure and cannot
exchange in course of chain fluctuations; the chemical structure is
prepared from "black" and "white" links at random with prescribed
probabilities.
\item {\it "Microcanonically quenched"} sequence---positions of "black"
and "white" links are again uniquely fixed for a given chain structure and
satisfy an extra constraint: the total number of "black" and "white" links
per chain is conserved for all realizations of chain structures.
\end{enumerate}

Our consideration is semi-analytical/semi-numerical. Namely, the closure of an
ideal $N$--link chain in a ring enables us to integrate from the very beginning
over all space degrees of freedom for quenched  sequences of links and
explicitly rewrite the Green's function of a chain as a quotient of
determinants of two $N\times N$--matrices with random coefficients on the main
diagonals. The transition point corresponds to the situation when the
denominator of the quotient becomes zero. This last question for large $N$ and
different models of chemical structures is analyzed numerically. 

The results are exposed in a form of histograms showing how many sequences have
the transition points at a given energy. Choosing the sequences which have the
transition points in the tail of the histogram we found that these sequences
have effective correlations between nearest--neighbor monomers for the model 
(2) and do not exhibit any correlations for the model (1).

\section{The model}

Consider an $N$--link ideal (i.e. self--intersecting) Gaussian polymer
chain in $d$--dimensional space. Suppose that the chain forms a ring and 
attach the first and the last chain's segments to the origin. All
segments are painted in "black" or "white" colors, designating different
interaction energies with the potential well located at the origin. There is
no interaction of chain's links between themselves or between a link and
any point in $d$--dimensional space distinct from the origin. Increasing
the interaction energy, we could provoke the transition from extended
(coil) to localized (adsorbed) state of the macromolecule. The transition
is governed by the interplay between entropic forces tending to expel the
chain and contact potential interactions which gain the energy in a compact
polymer configuration.

The said above can be easily rewritten in a formal way. Let us introduce
the joint distribution function $G_N({\bf x}_1,{\bf x}_2,...,{\bf x}_N)$
having sense of a probability that the links with numbers $1,2,...,N$ of
the $N$--step chain are located at the points ${\bf x}_1,{\bf x}_2,...,{\bf
x}_N$ in $d$--dimensional space. The Markov structure  of our polymer chain
ensures the presence of correlation only between the neighboring segments of
length $a$ and allows us to set
$$
P_N({\bf x}_1,{\bf x}_2,...,{\bf x}_N)=
g({\bf x}_1,{\bf x}_2)g({\bf x}_2,{\bf x}_3)...g({\bf x}_{N-1},{\bf x}_N)
$$
where the transition probability $g({\bf x}_1,{\bf x}_2)$ satisfies the
normalization condition $\int g({\bf x},{\bf x}') d{\bf x}=1$ and for
simplicity is supposed to be Gaussian:
\be \label{eq:1}
g({\bf x})=\left(\frac{d}{2\pi a^2}\right)^{d/2}
e^{-\frac{d|{\bf x}|^2}{2 a^2}}
\ee
The Green's function $G({\bf x}_N,N)\equiv\int G_N({\bf x}_1,{\bf x}_2,...,
{\bf x}_N)d{\bf x}_1 d{\bf x}_2...d{\bf x}_{N-1}$ is defined as follows
\be \label{eq:2}
G({\bf x}_N,N)=\int \prod_{i=1}^N \left[g({\bf x}_{i-1},{\bf x}_i)
e^{-\phi({\bf x}_i)/T}\right] d^dx_1...d^dx_{N-1}.
\ee
where $e^{-\phi({\bf x}_i)/T}$ is the Boltzmann weight of $i^{\rm th}$
chain segment in the potential well at the origin (compare to \cite{gro2}):
\be \label{eq:3}
e^{-\phi({\bf x}_i)/T}=1+\beta_i\delta({\bf x}_i); \quad
\beta_i=\left\{\begin{array}{ll}
e^{-u_{\rm b}}-1 & $\mbox{if the segment $i$ is "black"}$ \\
e^{-u_{\rm w}}-1 & $\mbox{if the segment $i$ is "white"}$
\end{array} \right.
\ee
and $u_{\rm b,w}$ is the dimensionless energy (everywhere below we set
$T=1$).

Now we can write the recursion equation for the Green's function
(\ref{eq:2}), expressing $G({\bf x}_{i+1},{i+1})$ in terms of $G({\bf x}_i,
i)$ and corresponding Boltzman weights for $0\le i\le N$. Performing
Fourier transform and taking into account (\ref{eq:3}) we arrive at the
following integral equation in the momentum space
\be \label{eq:4}
G({\bf k},i+1) = g({\bf k})G({\bf k},i) + (2\pi)^{-d}
\beta_{i+1}\int g({\bf k}')G({\bf k}',i)d^dk'
\ee
where
\be \label{eq:5}
g({\bf k})=e^{-|{\bf k}|^2 a^2/2d}
\ee
In order to make the system of equations (\ref{eq:4}) closed and nonuniform
we should take into account the initial and boundary conditions. Let us
point out that for open chains the precise form of initial and boundary
conditions does not affect in the thermodynamic limit the phase 
transition point (see, for example, \cite{gh}). The same we expect also for 
closed chains. Suppose that the last segment (with $i=N$) is linked to the 
origin by some sufficiently weak dimensionless
potential $f$. Without the loss of generality we may chose $f$ in the
simplest form: $f({\bf k})\equiv f={\rm const}$. The presence of
the potential $f$ increases the probability $\tilde{G}(k,N)$ to find the
$N$'s chain link near the origin. If $|f|\ll 1$ we may set in the linear
approximation:
\be \label{eq:6a}
\tilde{G}({\bf k},i=N)=G({\bf k},i=N)+f
\ee
Using the fact that the chain is closed, we complete equations (\ref{eq:4})
by the following one:
\be \label{eq:6b}
G({\bf k},i=1)=g({\bf k})\tilde{G}({\bf k},i=N) + (2\pi)^{-d}\beta_1\int
g({\bf k}')\tilde{G}({\bf k},i=N)d^dk'
\ee
(as we shall see later $f$ does not enter the final expression for the 
transition point).

The system of $N$ equations (\ref{eq:4})--(\ref{eq:6b}) is closed. This
enables us to write it as a single {\it matrix--integral} equation for the
$N$--component "spinor" Green's function ${\bf G}({\bf k})=\{G({\bf k},1),
G({\bf k},2),..., G({\bf k},N-1), G({\bf k},N)\}$. Hence we get:
\be \label{eq:7}
{\bf G}({\bf k})= \hat{M}g({\bf k}){\bf G}({\bf k})+
(2\pi)^{-d}\hat\beta \hat{M}\int g({\bf k}'){\bf G}({\bf k}')d^dk'+
{\bf F}({\bf k})
\ee
with
\be \label{eq:8}
\hat{M}=\left(\begin{array}{ccccc}
0 & 0 & 0 & \dots & 1 \\
1 & 0 & 0 & \dots & 0 \\
0 & 1 & 0 & \dots & 0 \\
0 & 0 & 1 & \dots & 0 \\
\vdots & \vdots & \vdots &  & \vdots
\end{array}\right), \quad
\hat{\beta}=\left(\begin{array}{cccccc}
\beta_1 & 0 & 0 & \dots & 0 \\
0 & \beta_2 & 0 & \dots & 0 \\
0 & 0 & \beta_3 & \dots & 0 \\
\vdots & \vdots & \vdots & \ddots & \vdots \\
0 & 0 & 0 & \dots & \beta_N
\end{array}\right),
\ee
and
\be
{\bf F}({\bf k})=\Big(\left(g({\bf k})+(d/2\pi a^2)^{\frac{d}2}\right)f,0,
\dots,0\Big)^{T}
\ee

Let us rearrange the terms in (\ref{eq:7}) as follows
\be \label{eq:9}
{\bf G}({\bf k})= \left(\hat{I}-g({\bf k})\hat{M}\right)^{-1}
\Big[(2\pi)^{-d}\hat\beta \hat{M} \int g({\bf k}'){\bf G}({\bf k}') d^dk'
+ {\bf F}({\bf k})\Big]
\ee
where $\hat{I}$ is $N\times N$ identity matrix. Integrating l.h.s and r.h.s
of (\ref{eq:9}) with the weight $g({\bf k})$ over all ${\bf k}$, we arrive
at the {\it algebraic matrix} equation for the function ${\bf
A}=\{A_1,A_2,...,A_N\}$, where $A_i=\int g({\bf k})G({\bf k},i) d^dk$:
\be \label{eq:10}
{\bf A}=(2\pi)^{-d} \left[\int g({\bf k}) \left(\hat{I}-g({\bf
k})\hat{M}\right)^{-1}\hat{\beta} \hat{M}\; d^dk \right] {\bf A}+{\bf F}
\ee

The straightforward computations show that the $N\times N$ matrix 
$\hat{C}=\left(\hat{I}-g({\bf k}) \hat{M}\right)^{-1} \hat{\beta} \hat{M}$ 
has the form:
\be \label{eq:11}
\hat{C}=\left(\begin{array}{lllll}
\beta_2 c_1 & \beta_3 c_2 & \beta_4 c_3 & \dots & \beta_1 c_N \\
\beta_2 c_N & \beta_3 c_1 & \beta_4 c_2 & \dots & \beta_1 c_{N-1} \\
\beta_2 c_{N-1} & \beta_3 c_N & \beta_4 c_1 & \dots & \beta_1 c_{N-2} \\
\vdots & \vdots & \vdots & \ddots & \vdots \\
\beta_2 c_2 & \beta_3 c_3 & \beta_4 c_4 & \dots & \beta_1 c_1
\end{array}\right)
\ee
with the coefficients
\be \label{eq:12}
c_i\equiv c_i({\bf k})=-\frac{g^{N-i}({\bf k})}{g^N({\bf k})-1},
\quad i\in[1,N]
\ee
and the vector ${\bf F}=(F_1,F_2,\dots,F_N)^T$ has components
\be
F_i=\frac{f}{(2 \pi)^d} \int \left(g({\bf k})+(d/2\pi
a^2)^{\frac{d}2}\right) c_{N-i+1}({\bf k}) d^dk
\ee

Hence the vector ${\bf A}$ obeys the system of equations
\be \label{eq:13}
\left\{\begin{array}{llllllllllllll}
A_1 & = & \beta_2 a_1 A_1 & + & \beta_3 a_2 A_2 & +
& ... & + & \beta_N a_{N-1} A_{N-1} & + & \beta_1 a_N A_N & + & F_1\\
A_2 & = & \beta_2 a_N A_1 & + & \beta_3 a_1 A_2 & +
& ... & + & \beta_N a_{N-2} A_{N-1} & + & \beta_1 a_{N-1} A_N & + & F_2\\
A_3 & = & \beta_2 a_{N-1} A_1 & + & \beta_3 a_N A_2 & +
& ... & + & \beta_N a_{N-3} A_{N-1} & + & \beta_1 a_{N-2} A_N & + & F_3\\
 ... & & & & & & & & & \\
A_{N-1} & = & \beta_2 a_3 A_1 & + & \beta_3 a_4 A_2 & +
& ... & + & \beta_N a_1 A_{N-1} & + & \beta_1 a_2 A_N & + & F_{N-1}\\
A_N & = & \beta_2 a_2 A_1 & + & \beta_3 a_3 A_2 & +
& ... & + & \beta_N a_N A_{N-1} & + & \beta_1 a_1 A_N & + & F_N
\end{array} \right.
\ee
where the coefficients $a_i$ for $1\le i\le N$ are defined as follows
\be \label{eq:14}
a_i=(2\pi)^{-d}\int g({\bf k}) c_i({\bf k})d^dk =
\left(\frac{d}{2 \pi a^2 N}\right)^{d/2}\zeta\left(\frac{d}2,
\frac{N+1-i}{N}\right);
\ee
Solving the corresponding system of linear equations via the standard
Kramers method, we arrive at the following expression for the values
$\beta_{i+1}A_i$:
\be \label{eq:15}
\beta_{i+1} A_i=
\frac{{\rm det}\hat{A}'(i)}{{\rm det}\hat{A}},
\quad i\in[1,N]; \quad \mbox{(by definition $\beta_{N+1}=\beta_1$)}
\ee
where the $N\times N$ matrix $\hat{A}$ reads
\be \label{eq:16}
\hat{A}=\left(\begin{array}{llllll}
a_1-\beta_2^{-1} & a_2 & a_3 & \dots & a_{N-1} & a_N \\
a_N & a_1-\beta_3^{-1} & a_2 & \dots & a_{N-2} & a_{N-1}\\
a_{N-1} & a_N & a_1-\beta_4^{-1} & \dots & a_{N-3} & a_{N-2}\\
\vdots & \vdots & \vdots & \ddots & \vdots & \vdots\\
a_3 & a_4 & a_5 & \dots & a_1-\beta_N^{-1} & a_2\\
a_2 & a_3 & a_4 & \dots & a_N & a_1-\beta_1^{-1} \end{array}\right)
\ee
and the $N\times N$ matrix $\hat{A}'(i)$ is obtained from the
matrix $\hat{A}$ by replacing the $i^{\rm th}$ column by the
vector $-{\bf F}$.

Equations (\ref{eq:15})--(\ref{eq:16}) enable us to rewrite $N$ linearly
independent components of the "spinor" ${\bf G}({\bf k})=
\{G_1({\bf k}),G_2({\bf k}),..., G_N({\bf k})\}$ (see (\ref{eq:9})) in a
compact form
\be \label{eq:18}
G_i({\bf k})= (2 \pi)^{-d}\Big[{\rm det}\hat{A}\Big]^{-1}
\sum_{j=1}^N c_{j-i+1}({\bf k})\,{\rm det}\hat{A}'(j)
+(\hat I-g({\bf k})\hat M)^{-1}{\bf F}({\bf k})
\ee
where $c_{j-i+1}({\bf k})=c_{N-(j-i+1)}({\bf k})$ if
$j-i+1\le 0$.

The transition point from delocalized (coil) to the adsorbed (globule)
state of a heteropolymer ring is manifested in the divergence of the
Green's function $G_i({\bf k})$ (\ref{eq:18}) for any $i\in[1,N]$. Thus
the transition point is determined by the equation
\be \label{eq:19}
{\rm det}\hat{A}=0
\ee
where the matrix $\hat{A}$ is given by (\ref{eq:16}). Let us pay attention
to the fact that equation (\ref{eq:19}) is very general: it is valid for
any type of disorder and number of species (i.e. sorts of the links).

For any finite chain lengths we can define only a transition region which
due to the supposed self--averaging becomes sharper and sharper as the
chain length increases, tending to a single point in the thermodynamic
limit $N\to\infty$.

Before passing to numerical solution of (\ref{eq:19}) for different randomly
generated sequences belonging to canonically-- and microcanonically quenched
chemical structures, let us derive analytically the transition point for the
effective homopolymer ring chain. In this case one can preaverage the
partition function of the chain over the distribution of "black" and "white"
links. Let us take $u_{\rm b}=u,\;u_{\rm w}=-u$ and suppose that the number of
"black" links is equal to the number of "white" ones. Then
\be \label{eq:effhom}
\beta_i\equiv \beta_{\rm av}=\frac{e^{-u}+e^u}{2}-1 = \cosh u -1
\ee
for all $1\le i\le N$.

Remind that for $N$--link {\it open} homopolymer chain attached by one end
at the origin in a potential well, the critical value $\beta_{\rm open}$ of
a transition point in $d$--dimensional space in the limit $N\to\infty$
reads \cite{gro2,gin}
\be
\beta_{\rm open}^{-1}=\left(\frac{d}{2 \pi a^2}\right)^{d/2}
\left.\zeta\left(\frac{d}2\right)\right|_{{d=3}\atop {a=1}} \approx 0.86188
\ee

The transition point of a {\it closed} chain can be analytically evaluated in
the effective homopolymer case (\ref{eq:effhom}). If all $\beta_i$ are equal,
then the matrix $\hat A$ (see eq.(\ref{eq:10})) is so--called {\it circulant} 
and its eigenvalues are \cite{mehta}
\be
\lambda_m=\sum_{j=1}^{N}a_j e^{i 2 \pi m (j-1)/N}, \quad m\in[0,N-1]
\ee
Eq. (\ref{eq:19}) is true when $\beta^{-1}$ equals to one of the eigenvalues
$\lambda_m$ that must be real. Thus
\be \label{eq:beta4}
\begin{array}{lll}
\beta^{-1}_{\rm av} & = & \disp \sum_{j=1}^{N}a_j=
\left(\frac{d}{2 \pi a^2 N}\right)^{d/2}\sum_{i=1}^N
\sum_{k=0}^{\infty} \frac1{(k+1+(1-i)/N)^{d/2}} \\
& = & \disp \left(\frac{d}{2 \pi a^2}\right)^{d/2}
\sum_{i=0}^{N-1} \sum_{k=1}^{\infty}
\frac1{(Nk-i)^{d/2}}=\left(\frac{d}{2 \pi a^2 }\right)^{d/2}
\sum_{j=1}^{\infty} \frac1{j^{d/2}} \\
& = & \disp \left(\frac{d}{2 \pi a^2}\right)^{d/2}
\zeta\left(\frac{d}2\right)=\beta_{\rm open}^{-1}
\end{array}
\ee
Hence, in 3D--space for $a=1$ we have 
$$
u_{\rm tr}^{\rm av}={\rm arccosh}[\beta_{\rm av}+1]=
{\rm arccosh}\left[\left(\frac{3}{2\pi}\right)^{3/2}
\zeta\left(\frac{3}{2}\right)+1\right]\approx 1.405
$$

Let us pay attention to the fact that the adsorption points of ring and open
chains coincide in the thermodynamic limit. This is consistent with the
statement on independence of the point of 2nd order phase transition on  the
boundary conditions \cite{ll}. Moreover, the interesting feature of
(\ref{eq:beta4}) consists in the fact that the value $\beta_{\rm av}$ does not
depend on $N$ for ring chains.

\section{Numerical results for quenched sequences of links.}

\begin{figure}[ht]
\begin{center}
\epsfig{file=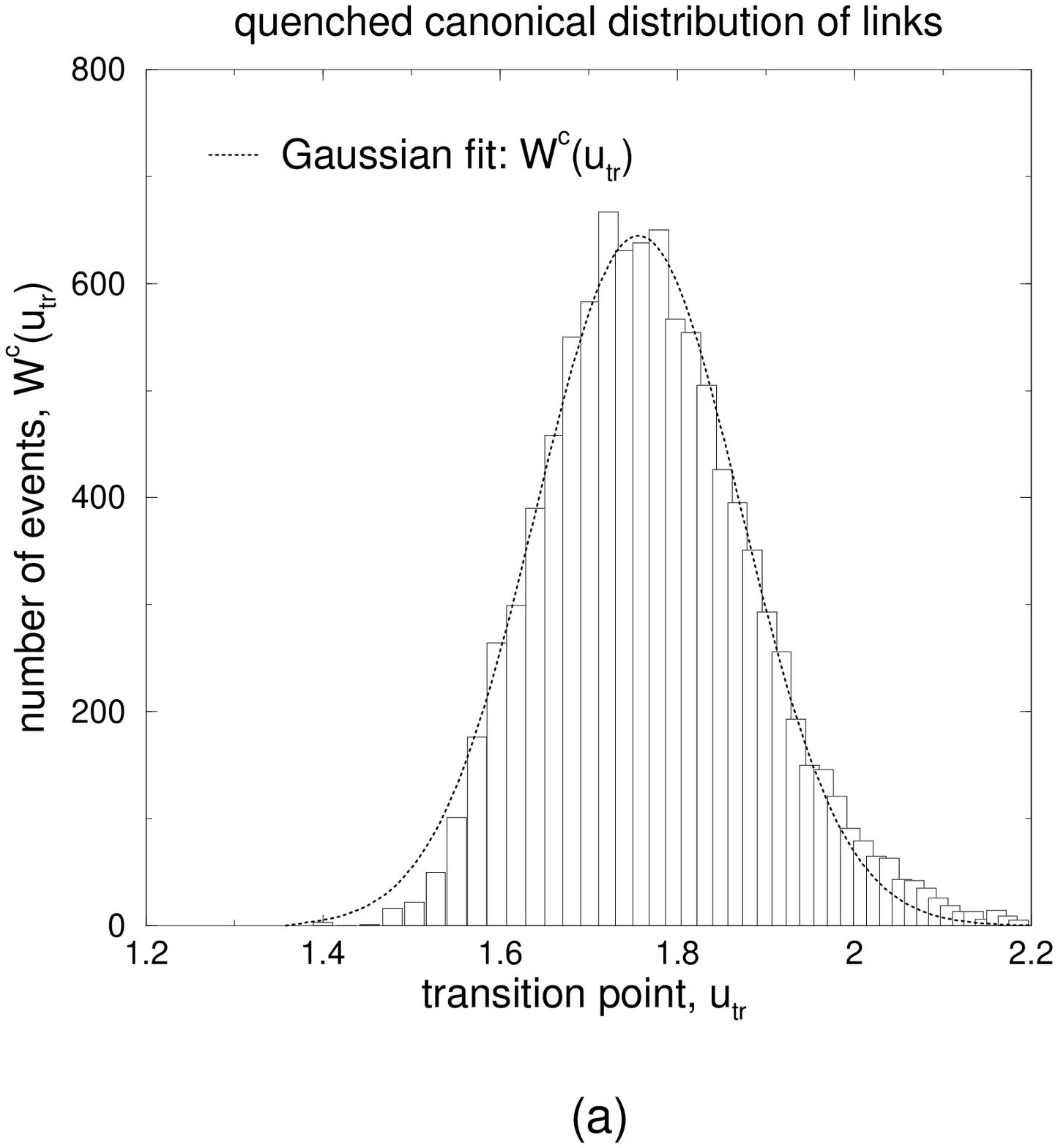,width=7.5cm}
\epsfig{file=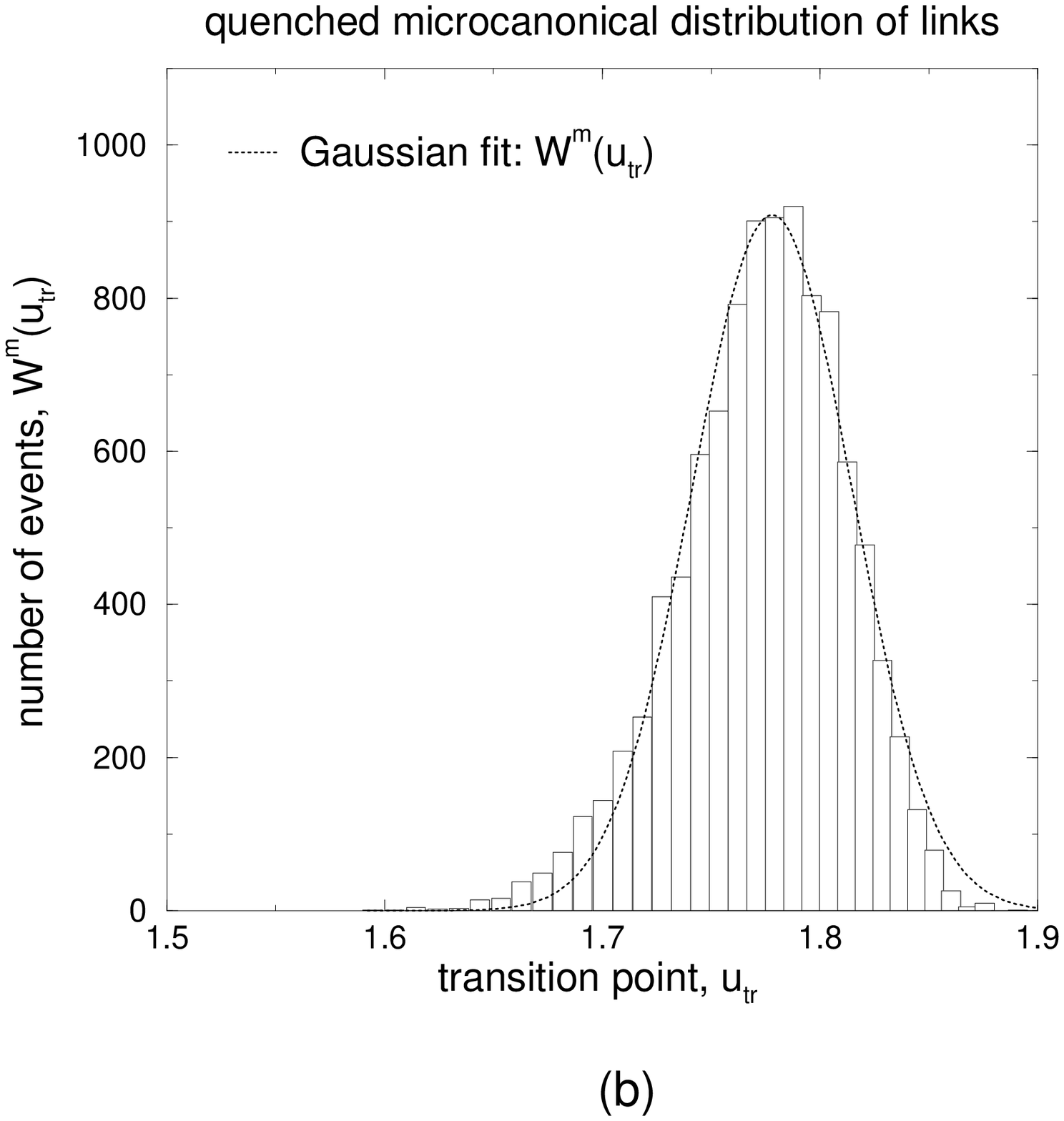,width=7.7cm}
\end{center}
\caption{Distribution of transition points for (a) canonical and (b)
microcanonical sequences of links.}
\label{fig:1}
\end{figure}

In this section we solve numerically Eq.(\ref{eq:19}) for two different types
of bimodal distribution of links with $u_{\rm b}=u,\;u_{\rm w}=-u$: (1)
canonically quenched and (2) microcanonically  quenched. Recall that in case
(1) "white" and "black" links appear in the sequence independently with equal
probability $\frac{1}{2}$; while in case (2) the total number of "black" and
"white" links per chain is exactly $\frac{N}{2}$. We have generated ensembles
of $10\;000$ chains of $N$ segments each for both of models (1) and (2). The
distribution of transition points $W(u)$ in cases (a) and (b) is depicted in
fig.\ref{fig:1} for $N=100$. 

The shape of histograms in fig.\ref{fig:1}a,b are correspondingly fitted by 
Gaussian curves
$$
W^{\rm c}(u_{\rm tr})=A^{\rm c} e^{-\frac{(u_{\rm tr}-
\left<u^{\rm c}_{\rm tr}\right>)^2}{2(\sigma^{\rm c})^2}}; \quad 
W^{\rm m}(u_{\rm tr})=A^{\rm m} e^{-\frac{(u_{\rm tr}-
\left<u^{\rm m}_{\rm tr}\right>)^2}{2(\sigma_{\rm m})^2}};
$$
For $N=100$ the parameters $A^{\rm c,m},\, \left<u^{\rm c,m}_{\rm tr}\right>,
\, \sigma_{\rm c,m}$ are as follows:
$$
\left\{\begin{array}{l}
A^{\rm c}=644.78 \\ \left<u^{\rm c}_{\rm tr}\right>=1.756 \\ 
\sigma_{\rm c}=0.116 \end{array}\right. \qquad
\left\{\begin{array}{l}
A^{\rm m}=908.49 \\ \left<u^{\rm m}_{\rm tr}\right>=1.778 \\ 
\sigma_{\rm m}=0.039 \end{array}\right.
$$

\begin{figure}[ht]
\begin{center}
\epsfig{file=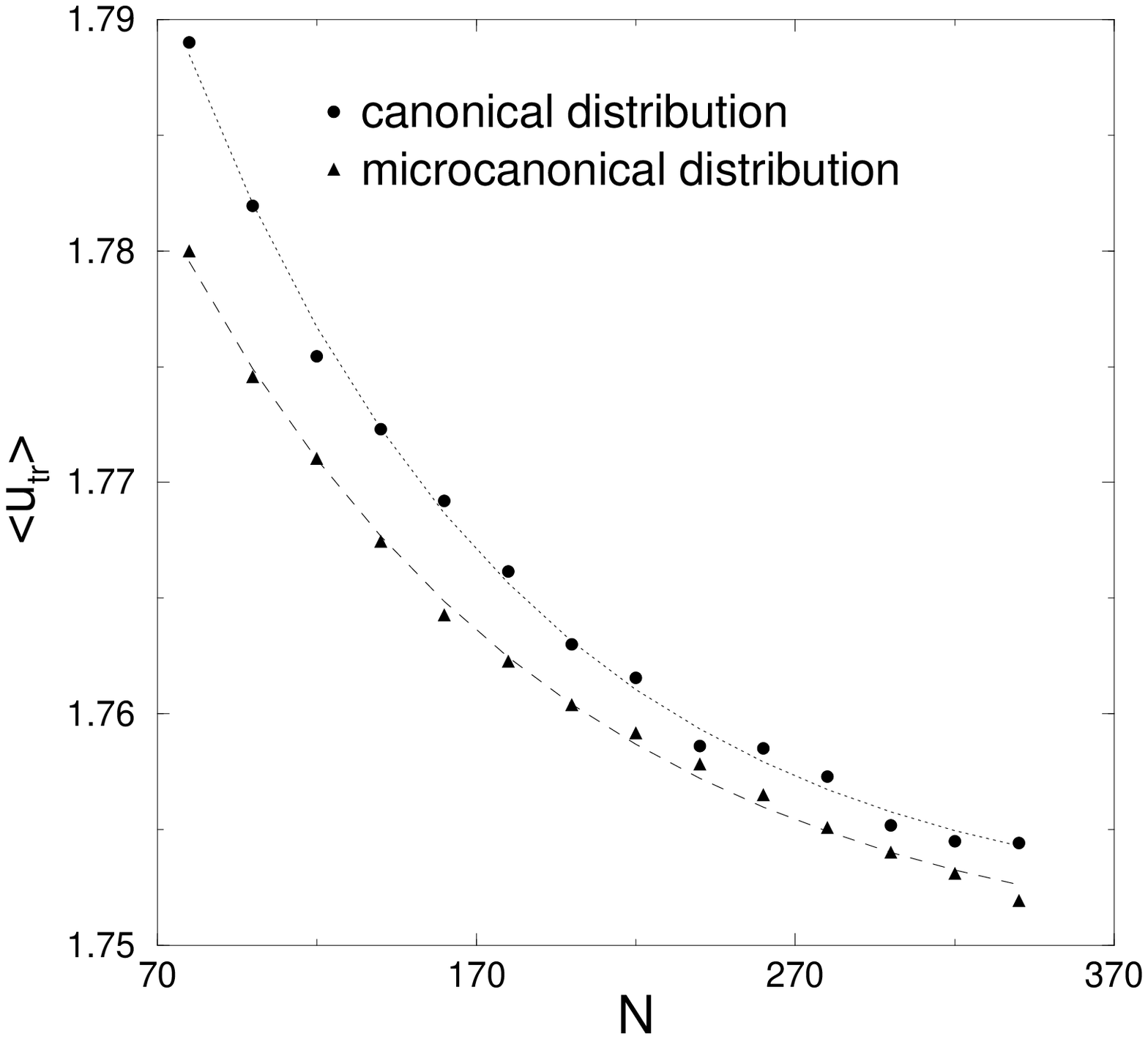,width=7cm}
\epsfig{file=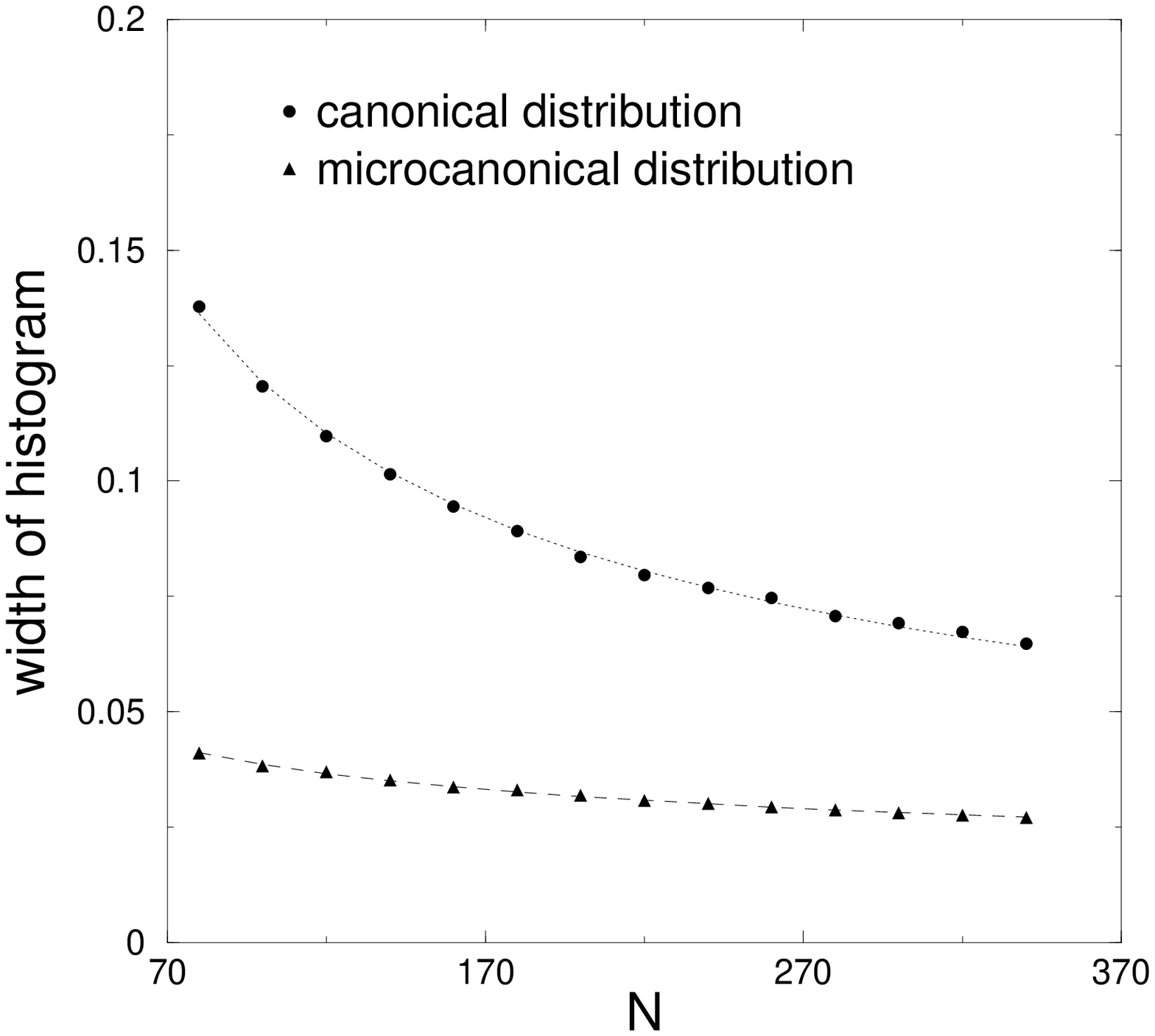,width=7cm}
\end{center}
\caption{Dependence of (a) mean value of transition point and (b) width of the
histogram on chain length.}
\label{fig:2}
\end{figure}

To predict the numerical values of $\left<u_{\rm tr}^{\rm c}\right>$ and  
$\left<u_{\rm tr}^{\rm m}\right>$ at $N\to\infty$ we have studied numerically 
the dependencies $\left<u^{\rm c}_{\rm tr}(N)\right>$ and 
$\left<u^{\rm m}_{\rm tr}(N)\right>$ for different $N$. The results of our 
simulations are displayed in fig.\ref{fig:2}a, where the chain lengths $N$ have
varied in the interval $80\le N\le 400$. The fitting curves are as follows
$$
\left<u^{\rm c}_{\rm tr}(N)\right>=u^{\rm c}_{\infty}+
u^{\rm c}_1 e^{-N/u^{\rm c}_2}; \quad
\left<u^{\rm m}_{\rm tr}(N)\right>=u^{\rm m}_{\infty}+
u^{\rm m}_1 e^{-N/u^{\rm m}_2}
$$
where
$$
\left\{\begin{array}{l}
u^{\rm c}_{\infty}=1.751 \\ u^{\rm c}_1=0.08 \\ u^{\rm c}_2=106.245
\end{array} \right. \qquad
\left\{\begin{array}{l}
u^{\rm m}_{\infty}=1.749 \\ u^{\rm m}_1=0.06 \\ u^{\rm m}_2=122.0.99
\end{array} \right.
$$

The crucial question concerns the behavior of the width of histograms (see
fig.\ref{fig:1}) in the thermodynamic limit $N\to\infty$. We have analyzed
numerically the behavior $\sigma^{\rm c,m}(N)$ for chains of length $N$ in the
interval  $80\le N\le 360$ and found no evident saturation of $\sigma^{\rm
c,m}$ at $N\gg 1$. The numerical data and corresponding power--law fits are
shown in fig.\ref{fig:2}b, where
$$
\sigma^{\rm c}(N)=\sigma^{\rm c}_0\,N^{-\alpha^{\rm c}} \qquad
\sigma^{\rm m}(N)=\sigma^{\rm m}_0\,N^{-\alpha^{\rm m}}
$$
and the numerical values are as follows:
$$
\left\{\begin{array}{lll}
\sigma^{\rm c}_0=1.331 &\quad \alpha^{\rm c}=0.52 & \mbox{for canoncal
distribution} \\
\sigma^{\rm m}_0=0.143 &\quad \alpha^{\rm m}=0.285 & \mbox{for microcanoncal
distribution}
\end{array}\right.
$$

\section{Discussion}

\subsection{Location of transition points in thermodynamic limit}

The results of our numerical simulations of eq.(\ref{eq:19}) permit us to
conclude that within the standard error the mean values of phase transition
energies  $\left<u_{\rm tr}^{\rm c}\right>$ and $\left<u_{\rm tr}^{\rm
m}\right>$ in  chains with symmetric bimodal canonically and microcanonically
quenched sequences  tend to one and the same value $\left<u_{\rm tr}^{\rm
c}\right>= \left<u_{\rm tr}^{\rm m}\right>\equiv u_{\infty} \approx 1.75\pm
0.02$ in the limit $N\to\infty$. At the same time the transition point in an
effective homopolymer chain defined by Eq.(\ref{eq:effhom}) is $u_{\rm tr}^{\rm
av}=1.405$. These numerical results seem to be sufficiently strong to 
conjecture the difference between the mean value of transition point for 
ensembles of heteropolymers with quenched sequences of links and the transition
point in an effective homopolymer chain. However one should honestly say that
the expressed conjecture deserves more investigation and still cannot be
considered as a rigorous statement. The reason why we cannot regard it yet as
an exact one is as follows. 

1. We have analyzed the distribution of transition points in sufficiently long
but finite chains ($N\sim 80\div 360$). Denote by $\Delta u_{\rm fin}(N)$
the transition region for finite $N$. As we know from the theory of 2nd order
phase transitions \cite{gh,ll}, 
\be \label{eq:concl1}
\Delta u_{\rm fin}(N)\sim N^{-1/2}
\ee
as $N\to\infty$ due to the {\it finiteness} of $N$ for homopolymer chains. 

2. The quenched randomness in types of links leads to an extra uncertainty in
the location of a transition point for heteropolymer chains. The transition
region $\Delta u_{\rm ran}(N)$ due to the {\it randomness} in sequences of
links  can be estimated by the width of the histograms---see fig.\ref{fig:1}.
So, we may set 
\be \label{eq:concl2}
\Delta u_{\rm ran}(N)\sim 
\left\{\begin{array}{ll} 
N^{-\alpha^c} & \mbox{for canonically quenched sequences} \\ 
N^{-\alpha^m} & \mbox{for microcanonically quenched sequences} 
\end{array}\right.
\ee

The behavior (\ref{eq:concl2}) is consistent with general self--averaging 
hypothesis for disordered systems \cite{mez} claiming the existence of unique
transition point for {\it every} disordered  sequence in the thermodynamic
limit $N\to\infty$, i.e.
$$
\Delta u_{\rm ran}(N)\Big|_{N\to\infty}\to 0
$$ 

Considering (\ref{eq:concl1}) and (\ref{eq:concl2}) as independent
contributions we arrive at the statement on the difference between 
$u_{\infty}$ and $u_{\rm tr}^{\rm av}$ in the limit $N\to\infty$. However one 
cannot exclude the possible "interference" between (\ref{eq:concl1}) and 
(\ref{eq:concl2}) which may lead to the smearing of the transition region. This
last possibility is not analyzed yet. Moreover, the asymptotics
(\ref{eq:concl2}) is based on the interpolation of numerical results for finite
$N$ to $N\to\infty$. The importance of (\ref{eq:concl2}) deserves an
analytic support of our conclusion.

\subsection{Design of sequences}

The data of our numerical simulations stored in course of the construction of 
the histograms (see fig.\ref{fig:1}) allow us to regard the following
problem. Let us define the correlation function $\chi$ for each ring chain
\be \label{eq:concl3}
\chi=\frac{1}{N}\sum_{j=1}^{N-1} \sigma_j\, \sigma_{j+1}
\ee
where the sum runs along all the chain's segments $j=[1,...,N]$ and 
$$
\sigma_j=
\left\{\begin{array}{ll}
+1 & \mbox{for "black" segment} \\
-1 & \mbox{for "white" segment}
\end{array}\right.
$$

\begin{figure}[ht]
\begin{center}
\epsfig{file=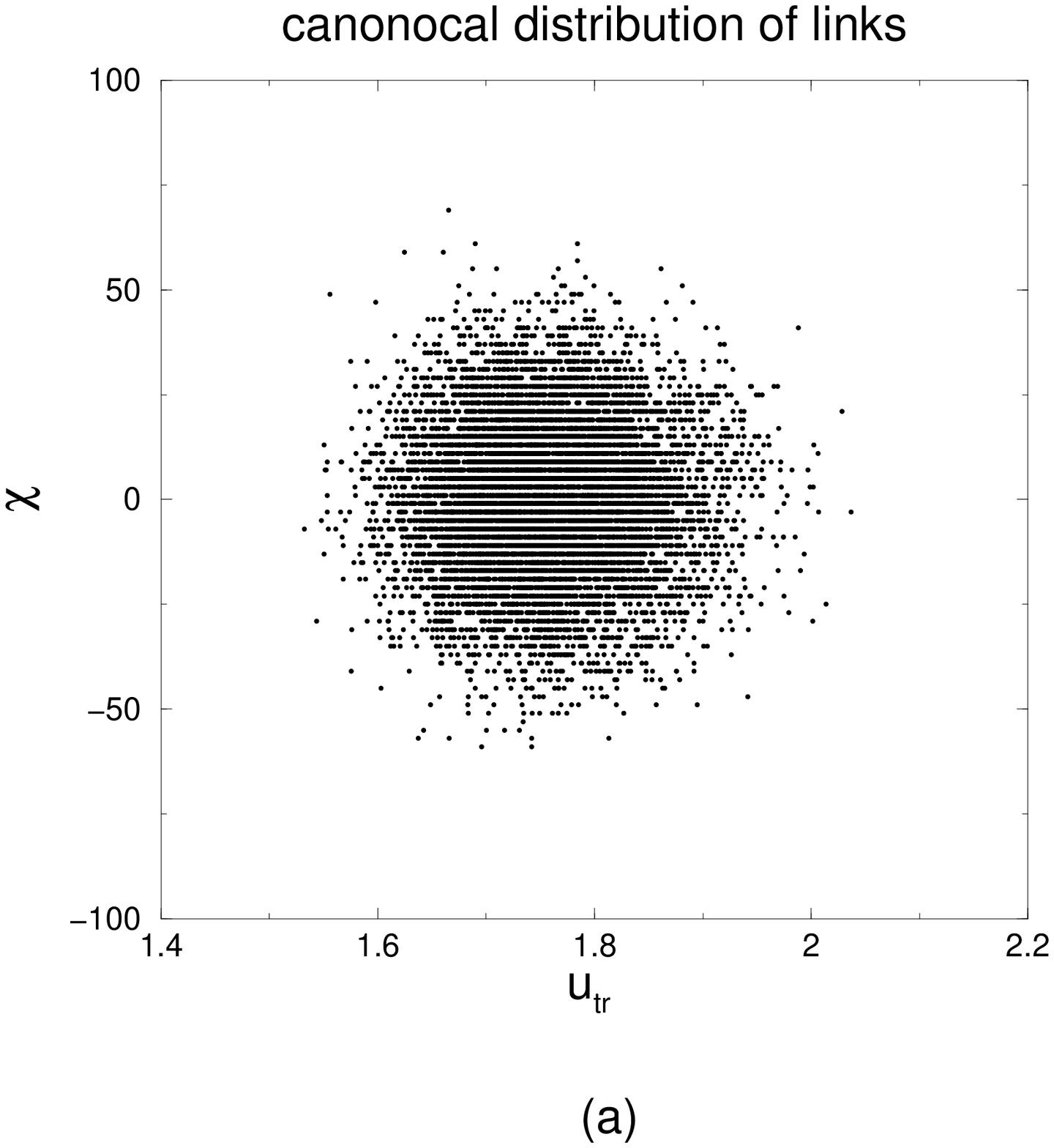,width=7.5cm}
\epsfig{file=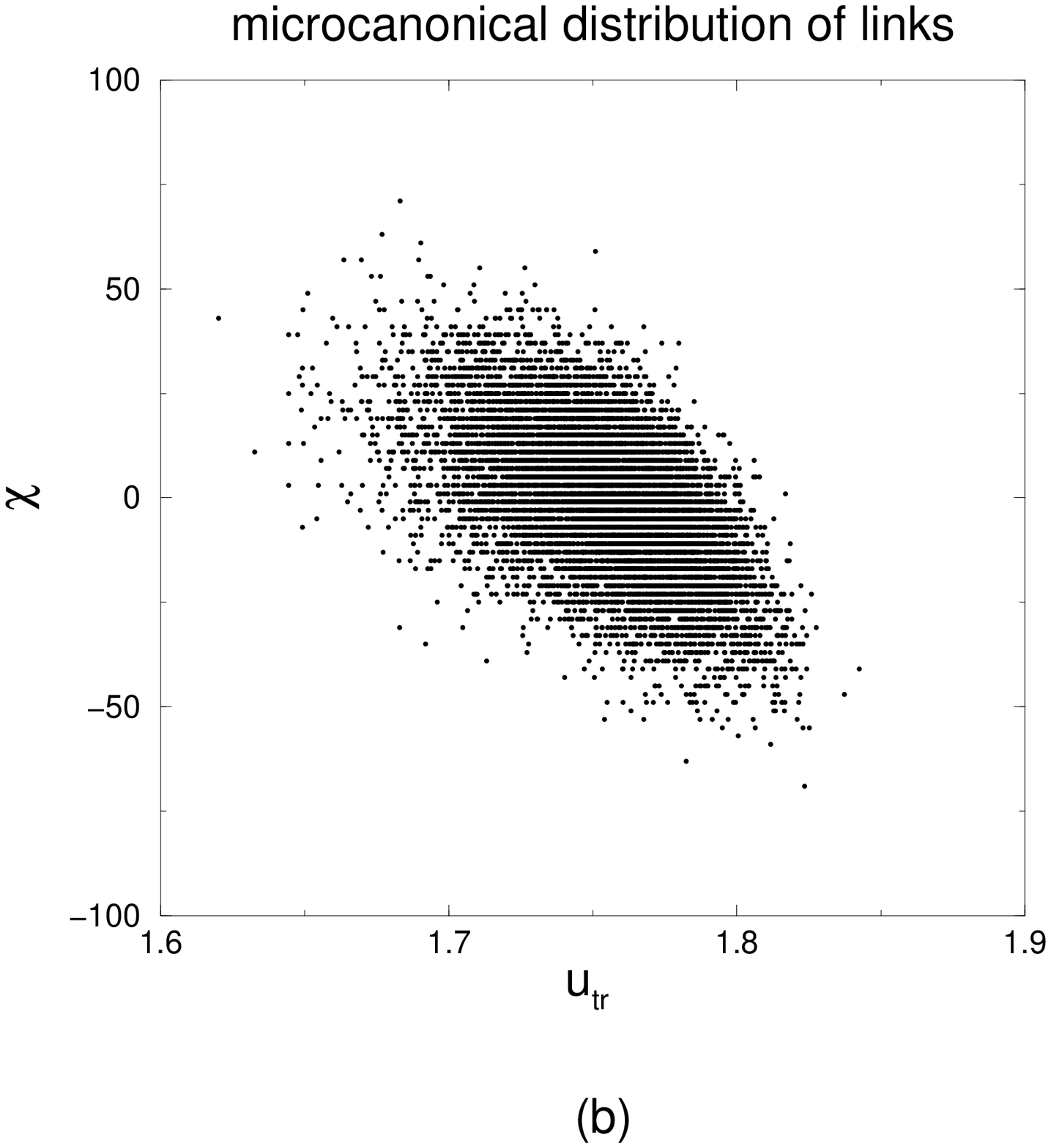,width=7.5cm}
\end{center}
\caption{The two--dimensional set $(u,\chi(u))$ for (a) canonically 
and (b) microcanonically quenched sequences; $N=300$.}
\label{fig:3}
\end{figure}

Knowing the transition point (the solution of Eq.(\ref{eq:19})) for every of
$10\;000$ sequences (for both of models (1) and (2)), let us assign to each 
sequence the point in the coordinates $(u,\chi(u))$ on the plane. The
corresponding "spots" for the models (1) and (2) are shown in
fig.\ref{fig:3}a,b for $N=300$. These figures are lead to the following 
important conclusions. 

1. There are no essential visible correlations between nearest--neighbor
links in the ensemble of canonically quenched sequences what is manifested in
the fact that the corresponding "spot" is circular---see fig.\ref{fig:3}a.

2. There are correlations between nearest-neighbor links in the ensemble of
microcanonically quenched sequences what is manifested in the fact that the
corresponding "spot" is asymmetric. Moreover, the figure \ref{fig:3}b allows to
conclude how the primary structure of the heteropolymer chain is organized: the
positive correlations $\left<\chi(u)\right> >0$ for $u<\left<u^{\rm m}\right>$ 
show the preferable clustering in the subsequences of the same color,
while the negative correlations $\left<\chi(u)\right> <0$ for $u>\left<u^{\rm
m}\right>$ show the preferable mixing of opposite colors.

Thus, knowing the value of the transition point for the chain of
"microcanonically quenched ensemble" we can make an approximate conclusion
about the primary sequence of monomers in a given chain. The fact that
the width of the histogram shrinks to zero at $N\to\infty$ means that 
apparently at $N\to\infty$ the corresponding "spot" in fig.\ref{fig:3}b becomes
more and more symmetric and the nontrivial correlations disappear in the
thermodynamic limit. However every real physical polymer system consists of
finite number of monomers what means that the above mentioned effect should 
in principle exist.

\bigskip

We expect that the approach proposed in the present work turns the initial
problem into well--posed subject of statistical theory of random matrices, what
would allow to apply the standard methods of random matrix theory for
heteropolymer adsorption.

\bigskip

\centerline{\bf Acknowledgments}

We are grateful to A.Grosberg for critical remarks and fruitful discussions and
to M.Mezard for useful suggestion.

\end{document}